\documentclass[fleqn,usenatbib]{mnras}
\usepackage[T1]{fontenc}

\usepackage{newtxtext,newtxmath}

\usepackage{ae,aecompl}
\usepackage{graphicx}
\usepackage{amsmath}
\usepackage{bm}	
\usepackage{hyperref}
\usepackage{xcolor}
\newcommand{\hMsun}{ h^{-1}{\rm M_{ \odot}}}
\newcommand{\hMpc}{ h^{-1}{\rm Mpc}}

\newcommand{\ihMpcC}{ h^{3}{\rm Mpc}^{-3}}

\newcommand{\hkpc}{ h^{-1}{\rm kpc}}
\newcommand{\ihMpc}{ h\,{\rm Mpc}^{-1}}

\newcommand{\vmax}{V_{\rm max}}
\newcommand{\vpeak}{V_{\rm peak}}
\newcommand{\mpeak}{M_{\rm peak}}
\newcommand{\minfall}{M_{\rm infall}}
\newcommand{\Fs}{f_{\rm s}}
\newcommand{\sigmaLogM}{\sigma_{\rm logM}}
\newcommand{\tmerger}{t_{\rm merger}}

\newcommand{\sig}{\sigma_{8}}
\newcommand{\OmM}{\Omega_\mathrm{M}}
\newcommand{\Omb}{\Omega_{\rm b}}
\newcommand{\h}{h}
\newcommand{\ns}{{n_{\rm s}}}
\newcommand{\Mnu}{M_{\rm \nu}}
\newcommand{\wa}{w_{\rm a}}
\newcommand{\wz}{w_{0}}

\title[The cosmological dependence of assembly bias]{The cosmological dependence of halo and galaxy assembly bias}

\author[S. Contreras et al.]{
S. Contreras,$^{1}$\thanks{E-mail: sergio.contreras@dipc.org}
J. Chaves-Montero$^{1}$,
M. Zennaro$^{1}$
 \&  R. E. Angulo$^{1,2}$.
\\
$^{1}$Donostia International Physics Center (DIPC), Manuel Lardizabal Ibilbidea, 4, 20018 Donostia, Gipuzkoa, Spain.\\
$^{2}$IKERBASQUE, Basque Foundation for Science, 48013, Bilbao, Spain.
}

\date{Accepted XXX. Received YYY; in original form ZZZ}

\pubyear{2021}

\begin{document}
\label{firstpage}
\pagerange{\pageref{firstpage}--\pageref{lastpage}}
\maketitle 

\begin{abstract}
One of the main predictions of excursion set theory is that the clustering of dark matter haloes only depends on halo mass. However, it has been long established that the clustering of haloes also depends on other properties, including formation time, concentration, and spin; this effect is commonly known as halo assembly bias. We use a suite of gravity-only simulations to study the dependence of halo assembly bias on cosmology; these simulations cover cosmological parameters spanning 10$\sigma$ around state-of-the-art best-fitting values, including standard extensions of the $\Lambda$CDM paradigm such as neutrino mass and dynamical dark energy. We find that the strength of halo assembly bias presents variations smaller than 0.05 dex across all cosmologies studied for concentration and spin selected haloes, letting us conclude that the dependence of halo assembly bias upon cosmology is negligible. We then study the dependence of galaxy assembly bias (i.e. the manifestation of halo assembly bias in galaxy clustering) on cosmology using subhalo abundance matching. We find that galaxy assembly bias also presents very small dependence upon cosmology ($\sim$ 2$\%$-4$\%$ of the total clustering); on the other hand, we find that the dependence of this signal on the galaxy formation parameters of our galaxy model is much stronger. Taken together, these results let us conclude that the dependence of halo and galaxy assembly bias on cosmology is practically negligible.
\end{abstract}


\section{Introduction}
\label{sec:Introduction}
In the $\Lambda$CDM model, dark matter haloes grow in a hierarchical way via smooth accretion and merging with other haloes \citep{PressSchechter:1974}. In the simplest approximation, the excursion set theory of hierarchical clustering \citep{Bond:1991} predicts that halo clustering does not depend on any property besides their mass. On the contrary, cosmological simulations predict that the clustering of haloes depends on multiple halo properties at fixed mass, including concentration, spin, and formation time; this effect is commonly known as halo assembly bias \citep{Sheth:2004, Gao:2005, Gao:2007}.

Halo assembly bias is defined as the differences in the clustering of haloes of the same mass but different secondary property (e.g., halo age, concentration, spin). This effect has been robustly measured in cosmological simulations (e.g., \citealt{Gao:2005, Gao:2007, Mao:2018}). However, its existence in the real universe is still under debate, with several studies finding hints of this signal (e.g., \citealt{Berlind:2006,Yang:2006,Cooper:2010,Wang:2013b,Lacerna:2014a,Lacerna:2014b,Hearin:2015,Miyatake:2016,Saito:2016,Obuljen:2020}) or indications of a lack of it (e.g., \citealt{Campbell:2015b,Zu:2016b,Zu:2017,Busch:2017,Sin:2017,Tinker:2017a,Lacerna:2017}). 

Most of the attempts of measuring halo assembly bias do not take into consideration that the cosmology they assume as fiducial could be, to some extent, incorrect. While the clustering and evolution of dark matter haloes have a strong dependence on cosmology, it is commonly assumed that this is not the case for halo assembly bias. Few works have discussed this assumption. For instance, \cite{Lazeyras:2021} shows that the halo assembly bias predicted by $N$-body simulations should not depend on the neutrino total mass parameter. These authors make similar claims for the dependence of halo assembly bias on $\sig$. In any case, despite halo assembly bias being a ubiquitous prediction of cosmological simulations, there is no systematic study about the dependence of this effect upon cosmology. Motivated by this, we consider a suite of cosmological gravity-only simulations spanning different cosmologies to study the dependence of the halo assembly bias upon changing 8 different cosmological parameters: $\sig$, $\OmM$, $\Omb$, $\ns$, $\h$, $\Mnu$, $\wz$ \& $\wa$ between $z=0$ and $z=3$. We find almost no dependence of the halo assembly bias signal with cosmology. We only find small variations on halo assembly bias when varying $\OmM$, and only at $z=0$ and for low halo masses.

Additionally, we also measure the dependence of galaxy assembly bias on cosmology. Galaxy assembly bias is the change in the galaxy clustering produced by the existence of halo assembly bias. This effect was first studied by \cite{Croton:2007}, who found that the galaxies of a semi-analytic model have a 2-point correlation function signal 10\%-20\% higher than a sample of galaxies populating haloes as a function of just halo mass. In practice, they use a ``shuffling'' technique, which we address in more detail in \S\ref{sec:GAB}. Recent works have shown that galaxy assembly bias is also present in hydrodynamical simulations (e.g., \citealt{Artale:2018}) and subhalo abundance matching (e.g., \citealt{ChavesMontero:2016}), and its amplitude is different for varies across different models, number densities, redshift and galaxy samples (e.g., \citealt{C19, C20b}).

While the effect of galaxy assembly bias is not large, ignoring this effect could affect the analysis of galaxy clustering, like the inference of cosmological parameters (e.g., \citealt{Zentner:2014}). This is why there have been several attempts at measuring the galaxy assembly bias signal in the real universe (e.g., \citealt{Zentner:2019,Salcedo:2021,Yuan:2021}),   but as for halo assembly bias, there is still not a consensus if galaxy assembly bias exists. In most of these studies, it is assumed that galaxy assembly bias does not depend on cosmology, but in our knowledge, this has never been properly studied before. Since galaxy assembly bias is the combination of a cosmological effect (i.e. halo assembly bias) and the particular way galaxies populate haloes in a specific galaxy formation model (i.e. occupancy variation, \citealt{Zehavi:2018}), even if halo assembly bias is independent of cosmology, galaxy assembly bias could still depend on cosmology. 

Motivated by this, we measure how the amplitude of the galaxy assembly bias signal depends on cosmology. We use a standard subhalo abundance matching technique (SHAM, \citealt{Conroy:2006, Reddick:2013, ChavesMontero:2016,Lehmann:2017, Dragomir:2018}) and a subhalo abundance matching extended technique (SHAMe, \citealt{C21a}) to populate our dark matter simulations with galaxies. Similarly to our results for halo assembly bias, we find that galaxy assembly bias has a weak dependence on cosmology, and only for $\sig$ and $\wz$.

The layout of the paper is as follows. In \S 2 we introduce the simulations and galaxy population models used. In \S 3 we measure how the halo assembly bias signal depends on cosmology, and in \S 4 we measure how the galaxy assembly bias depends on cosmology. We finalize by summarizing our results in \S 6.

Unless otherwise stated, the units in this paper are $\hMsun$ for masses, $\hMpc$ for distances and km/s for the velocities. All logarithm values are in base 10. 

\section{Simulations and galaxy models}
\label{sec:Sims}

\subsection{Dark matter simulations}
\label{sec:dm_sims}
Our main suite of simulations covers 13 cosmologies, spanning a cosmological hypervolume roughly 10-$\sigma$ around the best fitting cosmology individuated by the Planck's collaboration\footnote{Our hypervolume is actually 10 $\sigma$ around Planck's bestfits for $\sig$, $\OmM$, $\Omb$ \& $\ns$. For $\h$ we use an even larger range to cover for SN measurements. For $\Mnu$ and $w_0$ we cover a region about $\sim5$ $\sigma$ around Planck. Finally, $\wa$ spans approximately $1\ \sigma$ around Planck's bestfit.} \citep{Planck:2018}. In particular here we extend the work of \cite{C20a} by showing the dependence of the halo assembly bias when changing 8 different cosmological parameters, namely $\sig,\ \OmM,\ \Omb,\ \ns,\ \h,\ \Mnu,\ \wz\ \&\ \wa$.

\begin{table}
    \centering
        \caption{Cosmological parameters of the three main cosmologies of the BACCO project, and the 13 pair simulations spanning a $10-\sigma$ parameter space around Planck's best fitting values used in this work. }
    \begin{tabular}{cccccccccc}
        \hline
        Cosmology &  $\OmM$ & $\Omb$ & $\h$ & $\ns$ & $\sig$ & $\Mnu$ [eV] & $\wz$ & $\wa$\\
        \hline
        Nenya  & 0.265 & 0.050 & 0.60 & 1.01 & 0.9 & 0 & -1 & 0 \\
        Narya  & 0.310 & 0.050 & 0.70 & 1.01 & 0.9 & 0 & -1 & 0 \\
        Vilya  & 0.210 & 0.060 & 0.65 & 0.92 & 0.9 & 0 & -1 & 0 \\
        \hline
        \multicolumn{9}{c}{Nenya cosmology, $\sig$ = 0.73, 0.9} \\
        \multicolumn{9}{c}{Nenya cosmology, $\OmM$ = 0.23, 0.4} \\
        \multicolumn{9}{c}{Nenya cosmology, $\Omb$ = 0.04, 0.06} \\
        \multicolumn{9}{c}{Nenya cosmology, $\ns$ = 0.92, 1.01} \\
        \multicolumn{9}{c}{Nenya cosmology, $\h$ = 0.6, 0.8} \\
        \multicolumn{9}{c}{Nenya cosmology, $\Mnu [\mathrm{eV}]$ = 0, 0.4} \\
        \multicolumn{9}{c}{Nenya cosmology, $\wz$ = -1.3, -0.7} \\
        \multicolumn{9}{c}{Nenya cosmology, $\wa$ = -0.3, 0.3} \\
        \hline 
    \end{tabular}
    \label{tab:cosmologies}
\end{table}

It is important to remark that this set of simulations was created to be used as a validation of the cosmology rescaling technique \citep{Angulo:2010,C20a}. For this reason, there are three main cosmologies, Narya, Nenya \& Vilya, and all other cosmologies are based on Nenya, changing one parameter at a time. In Tab. \ref{tab:cosmologies} we show the cosmological parameters of each of the simulations in our set. The simulations corresponding to the main cosmologies were run assuming a comoving cubical box of $L_{\rm box} = 512 \hMpc$ on a side. All the other simulations have slightly different box sizes, chosen to match the volume they would have if we were to rescale one of the main simulations to that cosmology. All simulations follow the evolution of $1536^3$ dark matter particles, whose initial positions and velocities have been set at $z=49$ using second-order Lagrangian perturbation theory. The corresponding particle mass is slightly different in each cosmology depending on the simulation volume, but always of the order of $m_{\rm p} \sim 3 \times 10^9 h^{-1}M_\odot$.

Each simulation was carried out with an updated version of {\tt L-Gadget3} \citep{Angulo:2012}, which is a lean version of {\tt GADGET} \citep{Springel:2005}, used to run the Millennium XXL simulation and the Bacco Simulations \citep{Angulo:2020}. We assumed a Plummer-equivalent softening length of $5 \hkpc$ to compute gravitational forces, with numerical parameters chosen so that the matter power spectrum exhibits a $2\%$-level convergence at $k<10\ihMpc$ at low redshift. Moreover, for each cosmology, we ran two simulations with same fixed initial density amplitudes but inverted phases, using the ``Fixed \& Paired'' technique, which allows us to suppress cosmic variance by at least 2 orders of magnitude on scales $k < 0.1 \ihMpc$ \cite{Angulo:2016}. An additional feature included in {\tt L-Gadget3} is the ability to store a diluted sample of dark matter particles. This sample is constructed by uniformly selecting 1 every $4^3$ particles in Lagrangian space and we use it to compute the bias of dark matter haloes.

This version of {\tt L-Gadget3} allows for an on-the-fly identification of haloes and subhaloes using a Friend-of-Friend algorithm \citep[{\texttt{FOF},}][]{Davis:1985} and an extended version of {\tt SUBFIND} \citep{Springel:2001}, respectively. Our updated version of {\tt SUBFIND} can efficiently compute on-the-fly properties that are non-local in time such as the peak subhalo mass ($\mpeak$), peak maximum circular velocity ($\rm \vpeak$), infall subhalo mass ($\minfall$), and mass accretion rate among others. In this work, we consider snapshots of haloes and subhaloes at redshifts $0 < z < 3$.

Finally, in order to test the robustness of our results, we use two additional sets of simulations. In particular, to improve the statistical significance of our findings, we consider a version of the simulations of the three main cosmologies with the same mass resolution but $\sim 10$ times larger volume, namely the BACCO simulations presented in \cite{Angulo:2020}, with $L_{\rm box} = 1440\ \hMpc$ and $N_{\rm part} = 4320^3$. Finally, to check the soundness of our results against different mass resolutions, we also use a version of the three main cosmology simulations with $L_{\rm box} = 512 \hMpc$ and $N_{\rm part} = 2288^3$, which is equivalent to 3.3 times higher mass resolution.

\subsection{Subhalo abundance matching extended}
\label{sec:SHAMe}
To study galaxy assembly bias in \S \ref{sec:GAB}, we populate our simulations using the {\bf S}ub-{\bf H}alo {\bf A}bundance {\bf M}atching {\bf e}xtended technique described in \cite{C21a}. This model start by populating the subhaloes of an N-Body simulation using a standard SubHalo Abundance Matching technique \citep[SHAM,][]{Conroy:2006,Reddick:2013,ChavesMontero:2016,Lehmann:2017,Dragomir:2018}. In its basic form, the SHAM matches the expected stellar mass of a galaxy population with a subhalo property. An additional scatter is added to this relation to better mimic observational and theoretical predictions. We use $\vpeak$ as our main subhalo property, which we define as the maximum circular velocity ($\vmax \equiv \max(\sqrt{GM(<r)/r})$) during the subhalo evolution. We match the stellar mass function of the TNG300 hydrodynamical simulation, which successfully reproduces the stellar mass function from observations \citep{Pillepich:2018}. We tested that the specific choice of the stellar mass function has a sub-percental impact on the galaxy clustering (in particular, the two-point correlation function) when choosing galaxies by a fixed number density. We also test some SHAMe model built using $\mpeak$ for some additional tests shown in Appendix~\ref{sec:appendix}.

After building the basic SHAM, the model adds orphans galaxies, which are satellite structures with a known progenitor that are not currently resolved by the simulation but are expected to still exist in the halo. Not only are orphans needed to reproduce the galaxy clustering of large number density samples, but \cite{Guo:2014} showed that they are also required when the target number density is low. We assume an orphan merges with its central structure only when the time since accretion exceeds a dynamical friction timescale, $t_{\rm infall} > t_{\rm dyn}$, with $t_{\rm dyn}$ the dynamical friction time computed using a modified version of the expression given by \citet[][Eq.7.26]{BT:1987}

\begin{equation}
t_{\rm dyn} = \dfrac{1.17\ \tmerger\ d_{\rm host}^2\ V_{\rm host} (M_{\rm host}/10^{13}\ h^{-1}\mathrm{M}_{\odot})^{1/2}}{G \ln(M_{\rm host}/M_{\rm sub}+1)\ M_{\rm sub}},
\end{equation}

\noindent where $\tmerger$ is a free parameter that effectively regulates the number of orphan galaxies; $\rm d_{host}$ is the distance of the subhalo to the centre of its host halo; $\rm v_{host}$ is the virial velocity of the host halo; $\rm M_{host}$ is the virial mass of the host halo, and $\rm M_{sub}$ is the subhalo mass. It is worth noticing that the model does not use orphan galaxies for $\tmerger=0$, and that it does not assume mergers between galaxies for $\tmerger=\infty$.

The next step is to remove substructures that have lost most of their subhalo mass.  Galaxies that live in subhaloes that have lost most of their subhalo have also lost part (or all) of their stellar mass via stripping processes \citep{Smith:2016}. To account for this effect, we follow \cite{Moster:2018} and assume that all the galaxies in subhaloes with current mass below a fraction $\Fs$ of their $\mpeak$ are disrupted,

\begin{equation}
M_{\rm sub} < \Fs\ M_{\rm peak}.
\end{equation}
\noindent where $\Fs$ is a free parameter in our model, and $\mpeak$ is the maximum mass ever attained by a subhalo during its evolution.

The SHAMe model is also capable of adding an adjustable level of galaxy assembly bias signal to the galaxy sample, following the procedure of \cite{C20b}. This feature was omitted during the particular implementation of this work, to focus on the intrinsic galaxy assembly bias signal of the model. For a more detailed discussion on this implementation, we refer the reader to \citealt{C21a}.

We built galaxy catalogues by selecting the most massive galaxies such that their abundance is equal to a number density of ${\rm n=0.01,\ 0.0316\ \&\ 0.001}$ $\ihMpcC$, equivalent to a minimum value of the stellar mass of $8.26\times 10^{9}, 2.93\times 10^{10}$ \& $ 6.46\times 10^{10}$ $\hMsun$. 

Following \cite{C20b}, the SHAMe parameters were chosen to fit the clustering (the projected correlation function and the multipoles of the correlation function) of the TNG300 hydrodynamical simulation \citep{TNGa, TNGb, TNGc, TNGd, TNGe} at a fixed number density of $\rm n=0.01$  $\ihMpcC$. The values of the parameters used are $\sigmaLogM =  0.39$, $\tmerger = 0.93$, $\Fs= 0.05$. We test other combinations of these parameters, even using a standard SHAM without scattering ($\sigmaLogM$ = 0, $\tmerger$ = 0 and $\Fs$ = 0) finding similar results as with our SHAMe model for low number densities, but without the possibility of going to higher number densities because of the limitations of the model.

\begin{figure}
\includegraphics[width=0.45\textwidth]{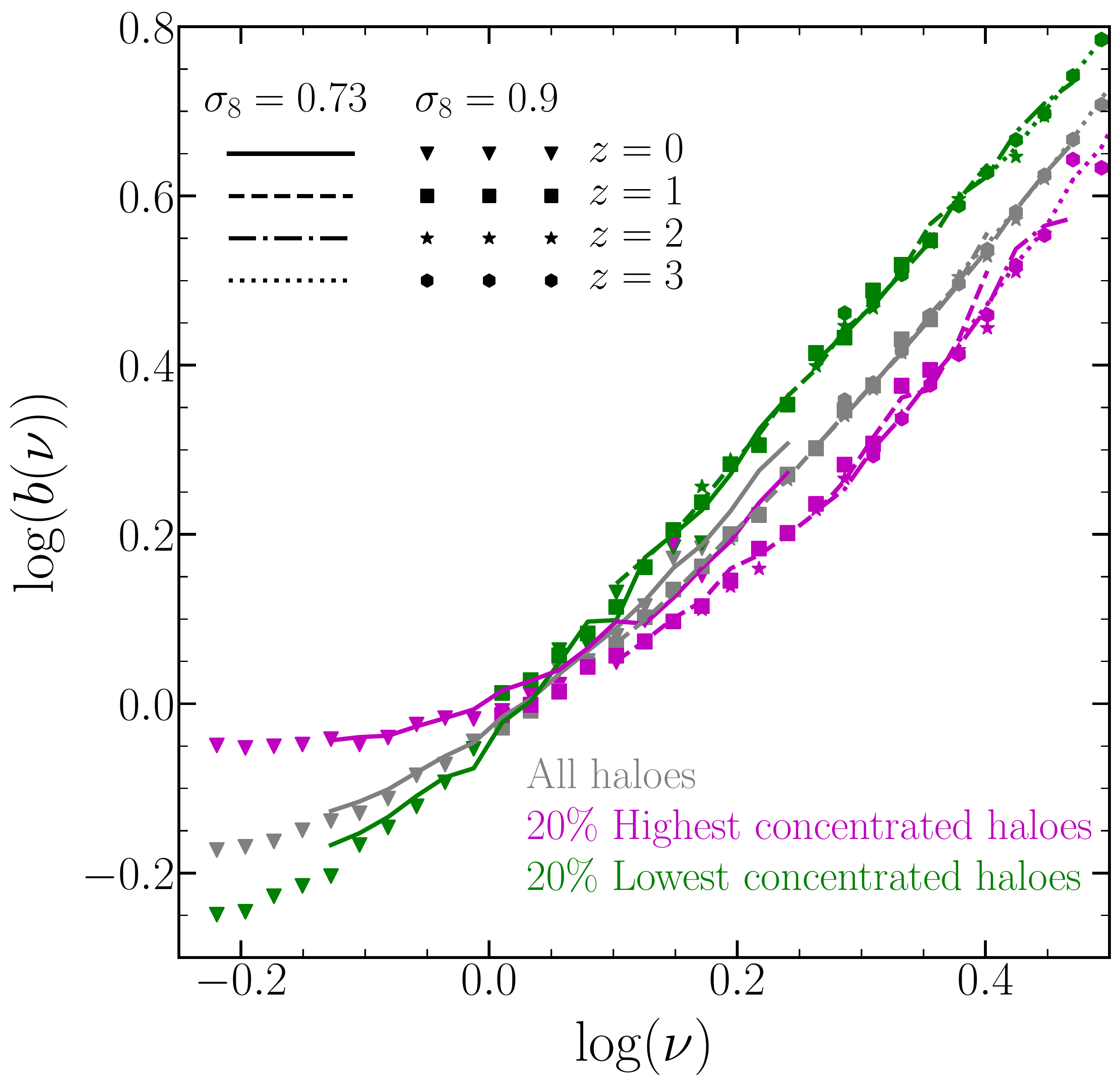}
\caption{
Dependence of halo assembly bias upon $\sig$ for haloes selected according to concentration. We show results extracted from two simulations using the same cosmology besides the amplitude of fluctuations, which is $\sigma_8=0.73$ for the first (lines) and $\sigma_8=0.90$ for the second (symbols). The grey colour indicates the linear bias of halos with different peak heights, green and magenta colours display the results for the 20\% of haloes with lowest and highest concentration at fixed peak height, respectively, and line and symbol styles depict the results at different redshifts. As we can see, the dependence of halo assembly bias on cosmology is negligible.
}
\label{Fig:hab_single}
\end{figure}

\begin{figure*}
\includegraphics[width=1.0\textwidth]{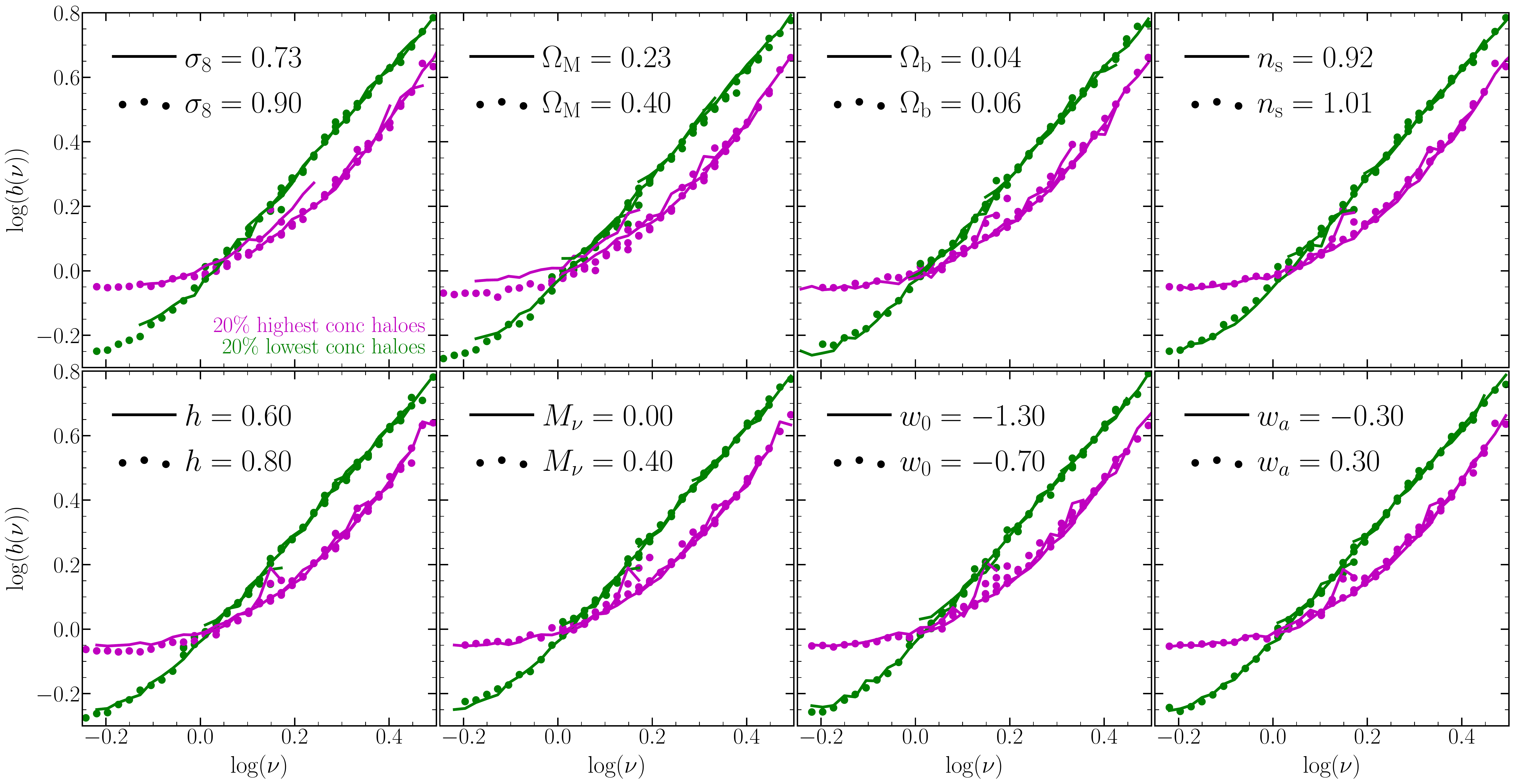}
\caption{
Dependence of halo assembly bias upon $\Lambda$CDM primary parameters, neutrino mass, and dark energy equation of state for haloes selected according to concentration. Each panel displays measurements from simulations varying the cosmological parameter indicated at the top right of the panel while holding fixed any other cosmological parameter to its fiducial value (see \S\ref{sec:dm_sims}). Green and magenta colours display the linear bias of the 20\% of haloes with the lowest and highest concentration at fixed peak height, respectively; lines and symbols indicate results from the simulation with the lowest and highest value of the cosmological parameter considered. As we can see, the linear bias of the most and lowest concentrated haloes is approximately the same for all cosmologies, letting us conclude that the dependence of halo assembly bias upon cosmology for haloes selected according to concentration is negligible.
}
\label{Fig:hab_conc}
\end{figure*}

\begin{figure*}
\includegraphics[width=1.0\textwidth]{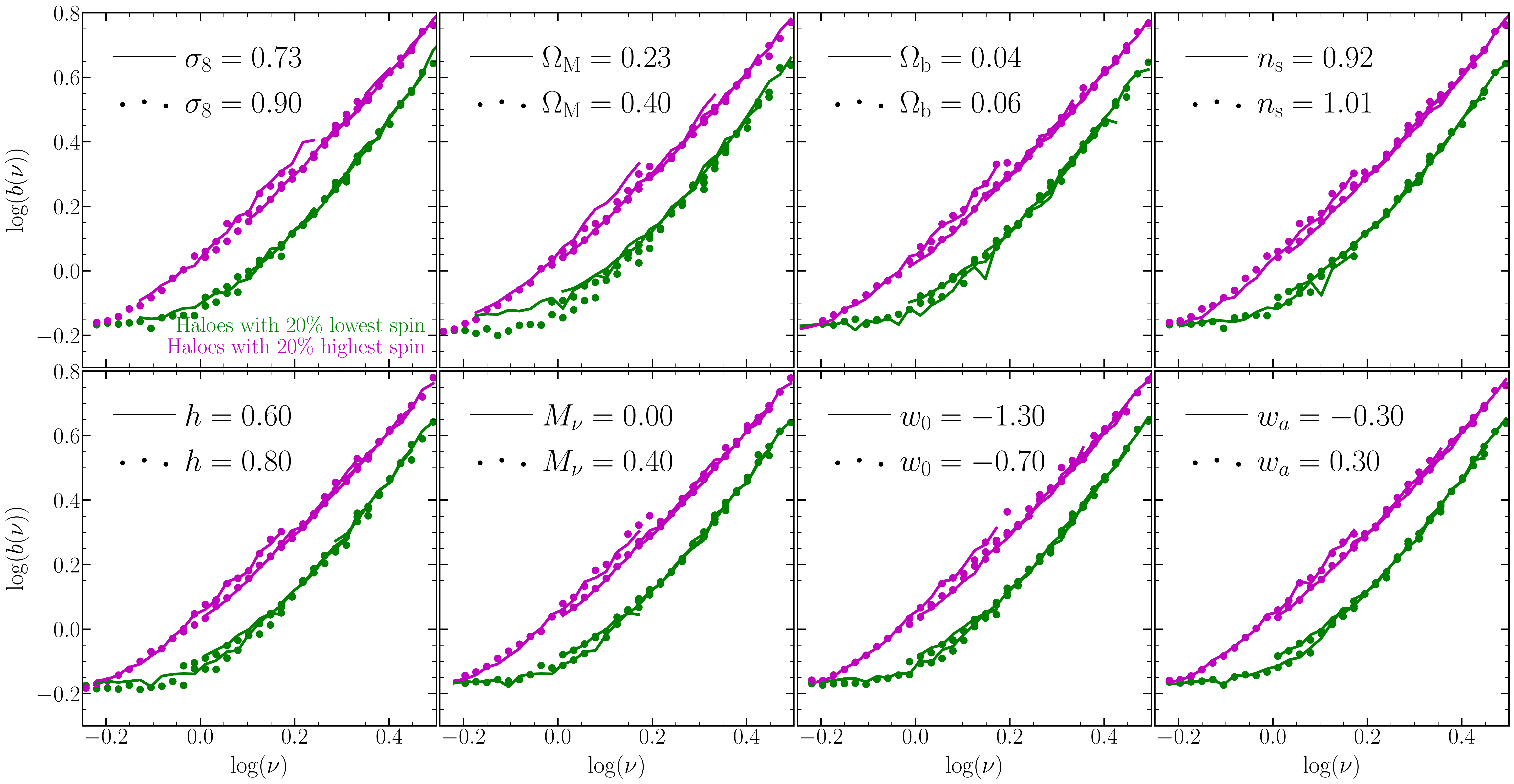}
\caption{Dependence of halo assembly bias upon cosmology for haloes selected according to spin. We use the same coding as in Fig.~\ref{Fig:hab_conc}. We find that the strength of halo assembly bias for spin-selected haloes is similar for all the cosmologies considered, in the same line as the results for concentration-selected haloes presented in Fig.~\ref{Fig:hab_conc}. Taken together, our findings suggest that halo assembly bias is cosmology independent.}
\label{Fig:hab_spin}
\end{figure*}

\begin{figure}
\includegraphics[width=0.45\textwidth]{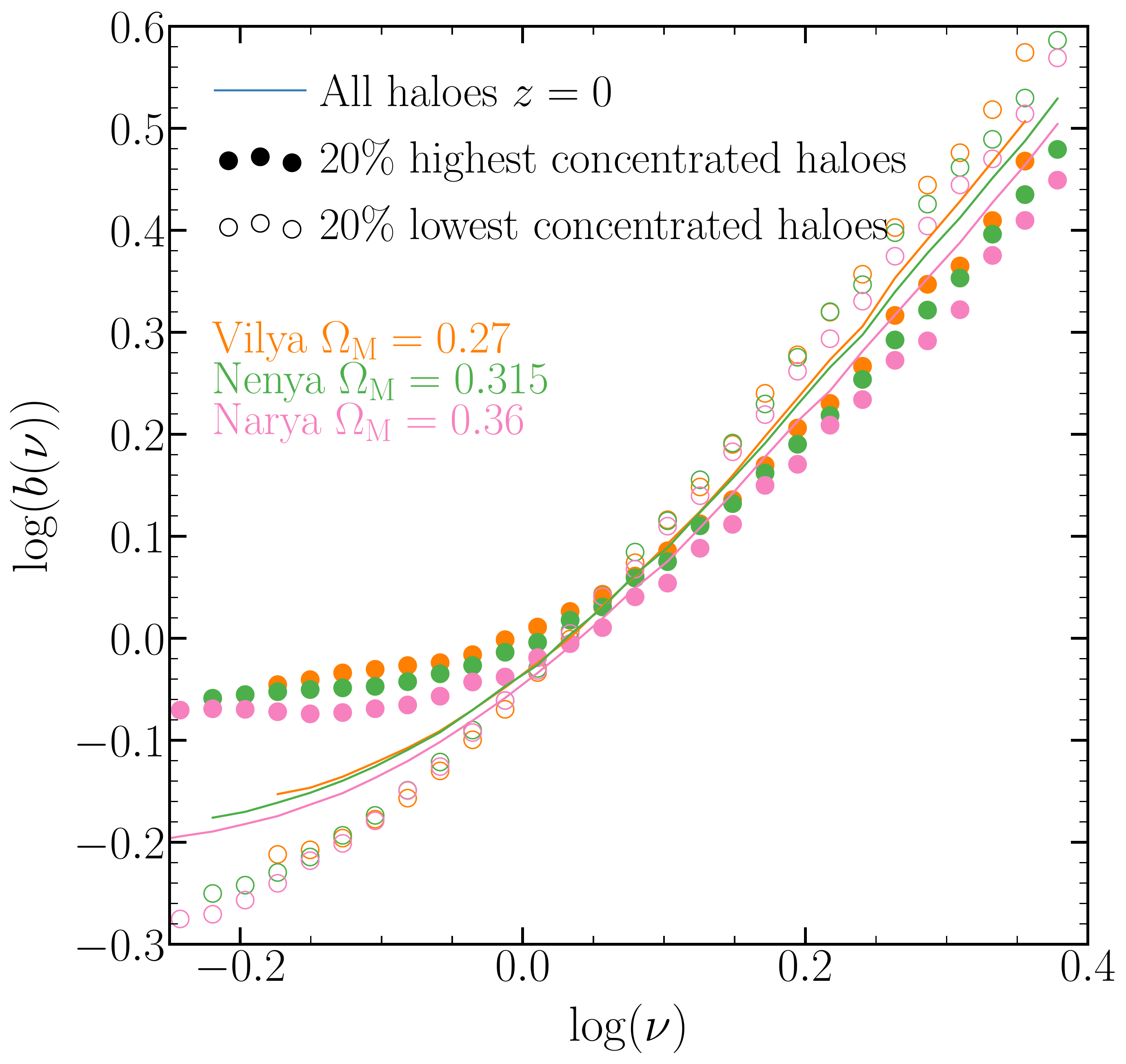}
\includegraphics[width=0.45\textwidth]{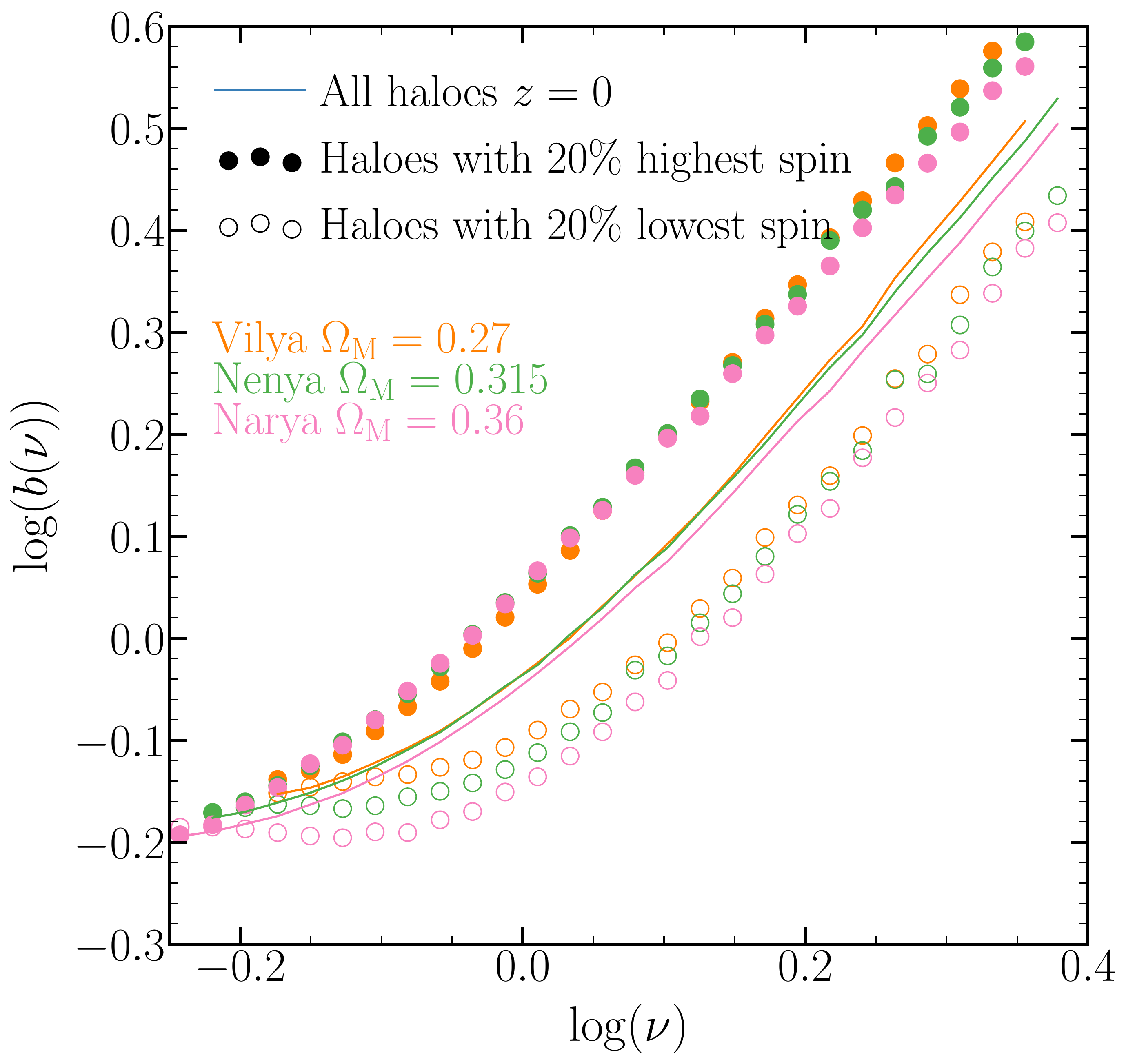}
\caption{Dependence of halo assembly bias on $\OmM$ at $z=0$. The top and bottom panels show the results for concentration- and spin-selected haloes, respectively, solid lines display the linear bias of haloes at fixed peak height, solid and open circles show the linear bias of the 20\% of haloes with the highest and lowest concentration or spin, and orange, green, and pink colours indicate measurements from $L_{\rm box} = 1440\,h^{-1}$ Gpc simulations with $\OmM=0.27,$ 0.315, and 0.36. We can readily see that the dependence of HAB on $\OmM$ for haloes with $\log\nu\lesssim0.05$ is weak but systematic: an increment  of $\Delta\Omega_\mathrm{M}=0.1$ induces an increase of linear bias of the order of 0.05 dex.}
\label{Fig:Rings_HAB}
\end{figure}

\section{Cosmology dependence of halo assembly bias}
\label{sec:HAB}

Halo assembly bias (HAB) refers to the dependence of the clustering of dark matter haloes on any internal halo property besides mass, including formation time, concentration, abundance of substructures, shape, and spin \citep{Sheth:2004, Gao:2005, Wechsler:2006, Gao:2007, Jing:2007, Faltenbacher:2010}. Interestingly, the strength of this effect varies across the properties mentioned above: for example, at $z=0$, halos with increasingly lower spin cluster more strongly at all masses, but more concentrated halos present stronger clustering than low concentrated haloes only at high masses \citep{Gao:2007}. Nonetheless, the dependence of HAB on halo properties has only been explored in detail at fixed cosmology; to our knowledge, only \cite{Lazeyras:2021} studied the dependence of this effect on neutrino mass. The primary purpose of this section is to study the dependence of HAB on each of the parameters of the $\Lambda$CDM model and some of its most common extensions.

To characterise the dependence of halo assembly bias upon cosmology, we select haloes according to peak height \citep{mo1996_AnalyticModelspatial}, $\nu(M, z) = \delta_c(z)/\sigma(M,\,z),$ where $\delta_c(z)$ is the linear overdensity threshold for collapse at redshift $z$ and $\sigma(M,\,z)$ refers to the variance of the linear overdensity field on a sphere containing a mass $M$ at redshift $z$. On average, haloes with peak height unity have just started to collapse, while those with peak height below or above unity have already collapsed or will collapse in the future \citep[e.g.,][]{mo1996_AnalyticModelspatial, sheth1999_LargescaleBiaspeak}. Notably, haloes selected according to peak height present similar properties independently of their mass and redshift \citep{sheth1999_LargescaleBiaspeak}.

We start our analysis by computing the peak height of halos in the $z=0,$ 1, 2, and 3 snapshots of each of the cosmological simulation described in \S\ref{sec:dm_sims}. Note that we run these simulations while holding fixed all cosmological parameters but one at a time, allowing us to study the dependence of halo assembly bias on each parameter separately. At each of these snapshots and for each simulation, we continue by selecting haloes according to their peak heights in bins of width 0.023 dex. To keep the level of statistic noise under control, we only consider peak height bins with more than 1000 haloes. We then draw the 20\% of haloes with the highest and lowest concentration from the first peak height bin at $z=0$, and we compute the cross-correlation between each of these two samples and the diluted dark matter density field (see \S\ref{sec:Sims}) at the same redshift using the publicly available {\sc corrfunc} package \citep[][]{Corrfunc1, Corrfunc2}. We extract the linear bias of each sample by computing the ratio between its cross-correlation with the diluted density field and the non-linear auto-correlation of the dark matter density field for the target cosmology. We predict the latter by applying the {\sc halofit} fitting function \citep{halofit} to the linear power spectrum of {\sc class} \citep{class}, and anti-transforming the resulting nonlinear power spectrum to configuration space. To ensure that the impact of non-linearities and cosmic variance is minimal, we compute the average ratio of the cross- and auto-correlations between $r=6$ and $20\,\hMpc$. Finally, we repeat the previous steps for all peak height bins at each redshift, and haloes selected according to concentration and spin. We check that the results remain the same when considering peak height bins of width twice and half the quoted value. Additionally, we extract linear biases from the BACCO simulations ($L_\mathrm{box}=1440\hMpc$) using larger scales (between $r=20$ and $65\,\hMpc$), finding analogous results as when using smaller scales.

Fig.~\ref{Fig:hab_single} shows the dependence of halo assembly bias upon $\sigma_8$ for haloes selected according to concentration. Lines and symbols display measurements from simulations using $\sigma_8=0.73$ and 0.90, respectively; note that these simulations present the same cosmology besides the value of $\sigma_8$. Magenta and green colours display the linear bias of the 20\% of haloes with highest and lowest concentration at fixed peak height, respectively, the grey colour shows the linear bias of haloes at fixed peak height with no selection over the concentration, and line and symbol styles indicate measurements from different redshifts. As naturally expected by the haloes' peak height selection, we find that the results do not depend on redshift. Motivated by this and to improve clarity, in subsequent plots we use the same line and symbol styles to show measurements at different redshifts. We can also see that grey symbols and grey lines agree precisely, and thus the dependence of the clustering of peak-height selected haloes on $\sigma_8$ is minimal. This finding is in agreement with the negligible dependence of the clustering of peak-height selected haloes on cosmology \citep{sheth1999_LargescaleBiaspeak}. 

The difference in the amplitude of the bias between the most and least concentrated haloes, i.e. the difference between magenta and green lines at fixed peak height, provides the strength of HAB for this property. As we can see, more concentrated halos present weaker clustering than less concentrated haloes for $\log\nu\gtrsim0.05$, while this trend reverses for $\log\nu\lesssim0.05$; therefore, HAB is positive (negative) for low concentrated haloes with peak height greater (smaller) than $\log\nu\simeq0.05$. These results are in the same line as the findings of \citet{Gao:2007}. The most noteworthy feature of Fig.~\ref{Fig:hab_single} is that the linear biases extracted from the simulations with $\sigma_8=0.73$ and 0.90 present approximately the same value at fixed peak height, letting us conclude that the dependence of HAB on $\sigma_8$ is minimal.

To check if the dependence of HAB on cosmology is also negligible for other cosmological parameters, in Fig.~\ref{Fig:hab_conc} we display linear biases extracted from pairs of simulations varying one cosmological parameter at a time. Therefore, this figure lets us assess the dependence of HAB on different cosmological parameters separately. Each panel shows measurements extracted from simulations varying the cosmological parameter indicated at the top left of the panel, lines and symbols display the results for the simulation with the lowest and highest value of the cosmological parameter under consideration, respectively, and magenta and green colours indicate the linear bias of the 20\% of haloes with highest and lowest concentration. As we can see, the linear bias of the highest and lowest concentrated haloes is approximately the same for all cosmologies; therefore, the dependence of HAB upon cosmology for concentration-selected haloes is minimal.

To further investigate the dependence of HAB on cosmology, we proceed to study the dependence of this effect on cosmology for spin-selected haloes. This is motivated by the difference dependence of HAB on concentration and spin, and the different physical origin of HAB for these two properties. Fig.~\ref{Fig:hab_spin} shows the dependence of halo assembly bias upon cosmology for haloes selected according to spin using the same coding as in Fig.~\ref{Fig:hab_conc}. As we can see, the strength of halo assembly bias for spin-selected haloes is similar for all the cosmologies considered, in the same line as the results for concentration-selected haloes. Taken together with the large variations in the cosmological parameters that we consider, in most cases equal or greater than $\pm 10\ \sigma$ around \cite{Planck:2018} values, our findings suggest that halo assembly bias is mostly cosmology independent.

In Figs.~\ref{Fig:hab_conc} and \ref{Fig:hab_spin}, it is possible to appreciate a slight dependence of HAB on $\OmM$ for $\log\nu\lesssim0.05$ haloes with high concentration and low spin; nonetheless, this dependence is so weak that it could be just due to cosmic variance. To reduce the impact of statistical uncertainties on the results, we consider three simulations with ten times more volume and $\OmM=0.27,$ 0.315, and 0.36 (see \ref{sec:dm_sims}). In Fig.~\ref{Fig:Rings_HAB}, we display measurements of HAB from these three simulations at $z=0$. The top and bottom panels show the results for concentration- and spin-selected haloes, respectively, solid lines display the linear bias of haloes at fixed peak height, solid and open circles show the linear bias of the 20\% of haloes with the highest and lowest concentration or spin, and orange, green, and pink colours indicate measurements from the simulations with $\OmM=0.27,$ 0.315, and 0.36. We can readily see that the same trends shown by Figs.~\ref{Fig:hab_conc} and \ref{Fig:hab_spin} emerges here: the linear bias of highly concentrated and low spin haloes with $\log\nu\lesssim0.05$ increases with $\OmM$. It is worth noting that this dependence is weak: variations of $\Delta\OmM=0.1$ induce changes of the order of 0.05 dex in linear bias.

As noted in \S\ref{sec:Sims}, the mass resolution of our simulations varies slightly with cosmology. To check whether the dependence of HAB on $\Omega_\mathrm{M}$ is the result of resolution-dependent biases in concentration and spin measurements, we consider cosmological simulations with 3.3 times higher mass resolution and the same box size, with $\Omega_\mathrm{M}=0.27,$ 0.315, and 0.36. Also in this case, we find the same trends. Moreover, we study whether this dependence is the manifestation of the different number density of backsplash haloes as a function of cosmology; these are current haloes that belonged to a higher mass halo in the past. We find the same HAB dependence on $\OmM$ after removing these objects. A further study of this difference, while interesting, is beyond the aim of this paper.

In conclusion, we find low-to-null dependence of the HAB signal on cosmology, with only some small departure for $\OmM$. While systematic, this signal induces linear bias variations smaller than 0.05 dex for extreme changes in cosmology and only for low mass haloes and low redshifts.

\begin{figure*}
\includegraphics[width=1.0\textwidth]{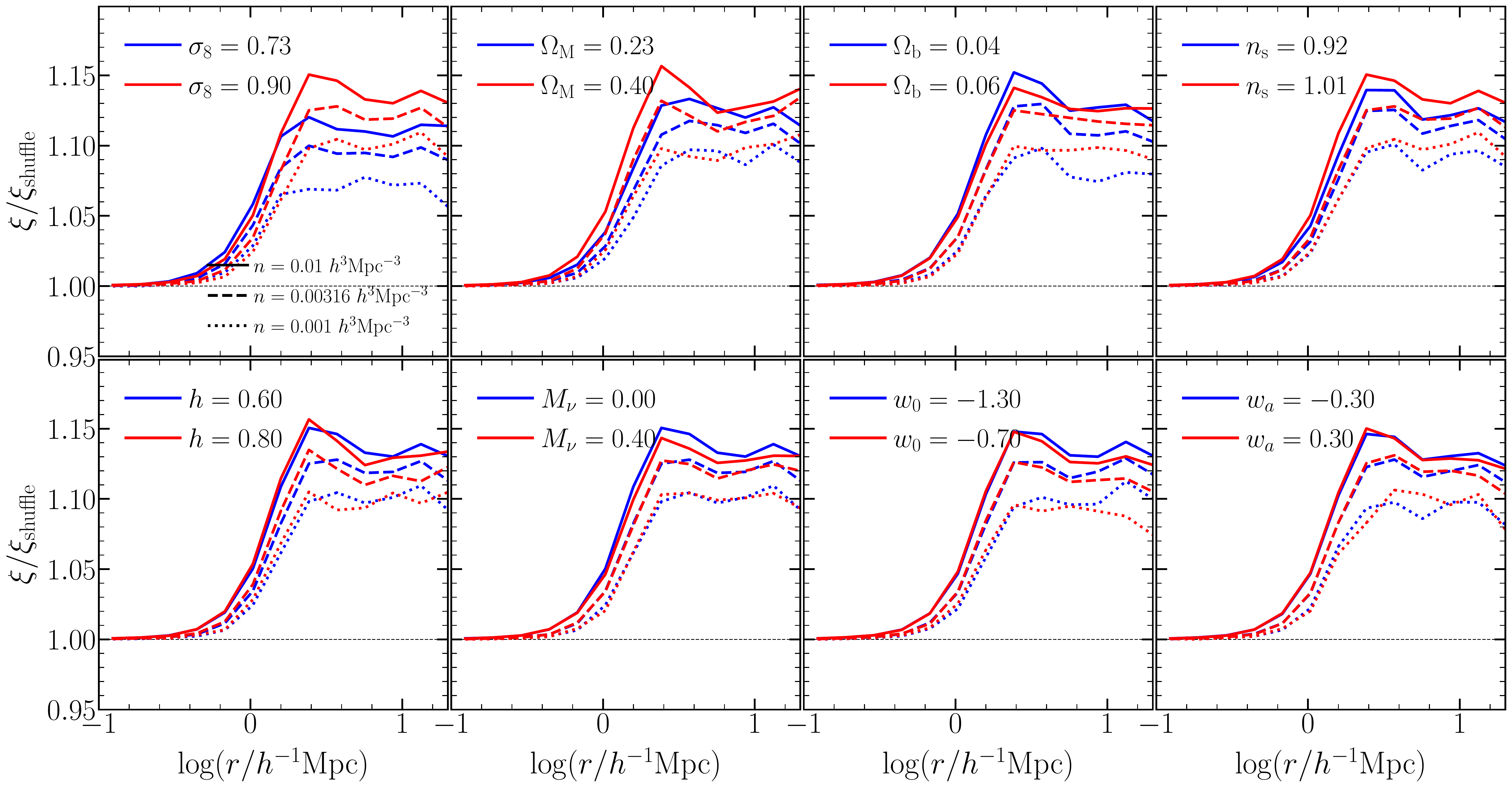}
\caption{The ratio of the 2-point correlation function of the galaxies of our extended SHAM with its shuffle counterpart at $z=0$(see \S~\ref{sec:GAB} for a full description of this method). The solid, dashed and dotted lines show the prediction for samples with three different number densities: n= 0.01, 0.00316 \& 0.001 $\ihMpcC$ respectively. Each panel shows results from two simulations varying one cosmological parameter (top row, from left to right: $\sig$, $\OmM$, $\Omb$ \& $\ns$; bottom row, from left to right: $\h$, $\Mnu$, $\wz$, $\wa$), with the lowest of these parameter show in blue and the highest in red.}
\label{Fig:gab_z0}
\end{figure*}
 
\begin{figure*}
\includegraphics[width=1.0\textwidth]{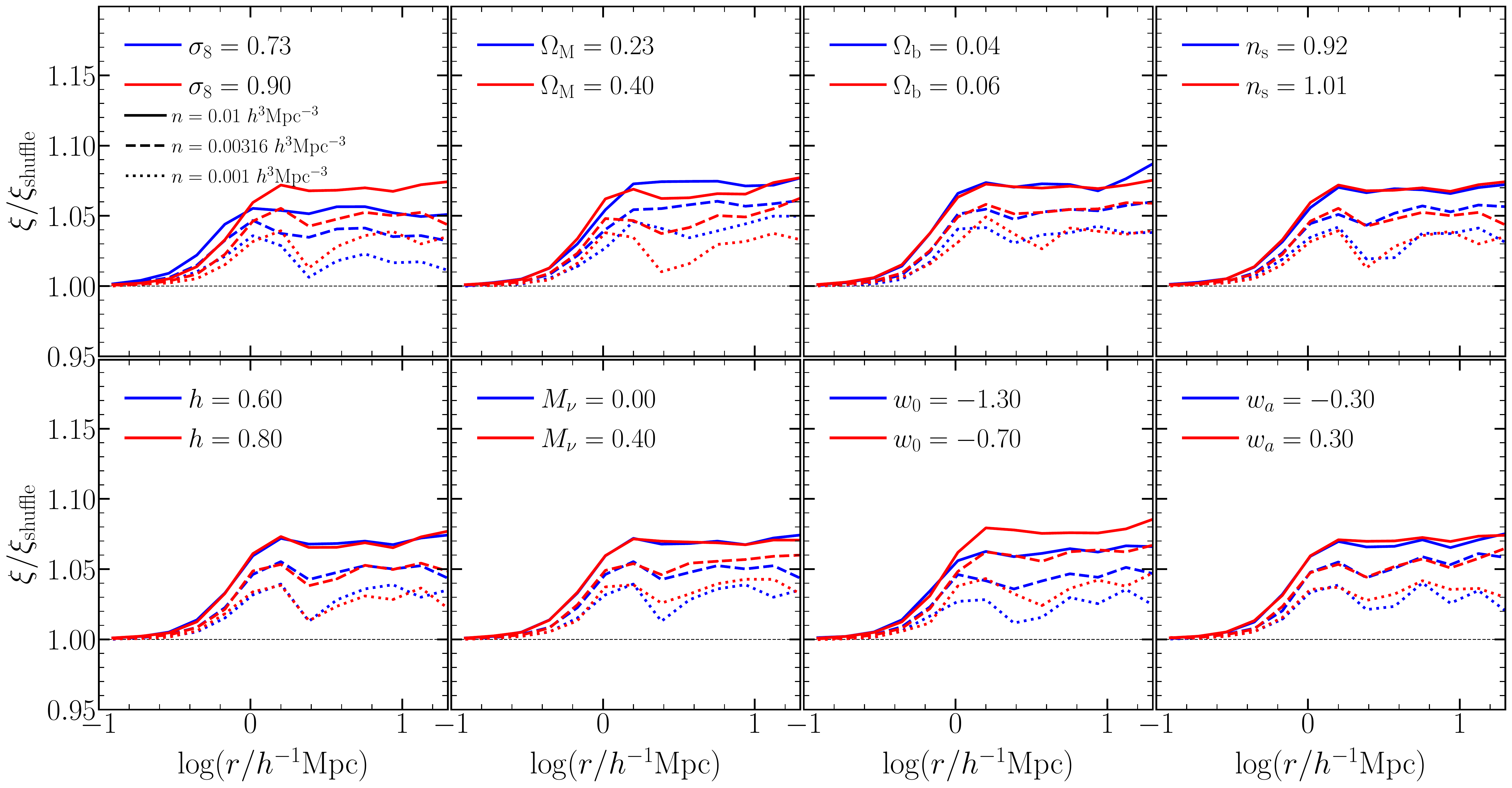}
\caption{Similar to Fig.~\ref{Fig:gab_z1}, but for galaxies at $z=1$.}
\label{Fig:gab_z1}
\end{figure*}
 
\begin{figure*}
\includegraphics[width=0.45\textwidth]{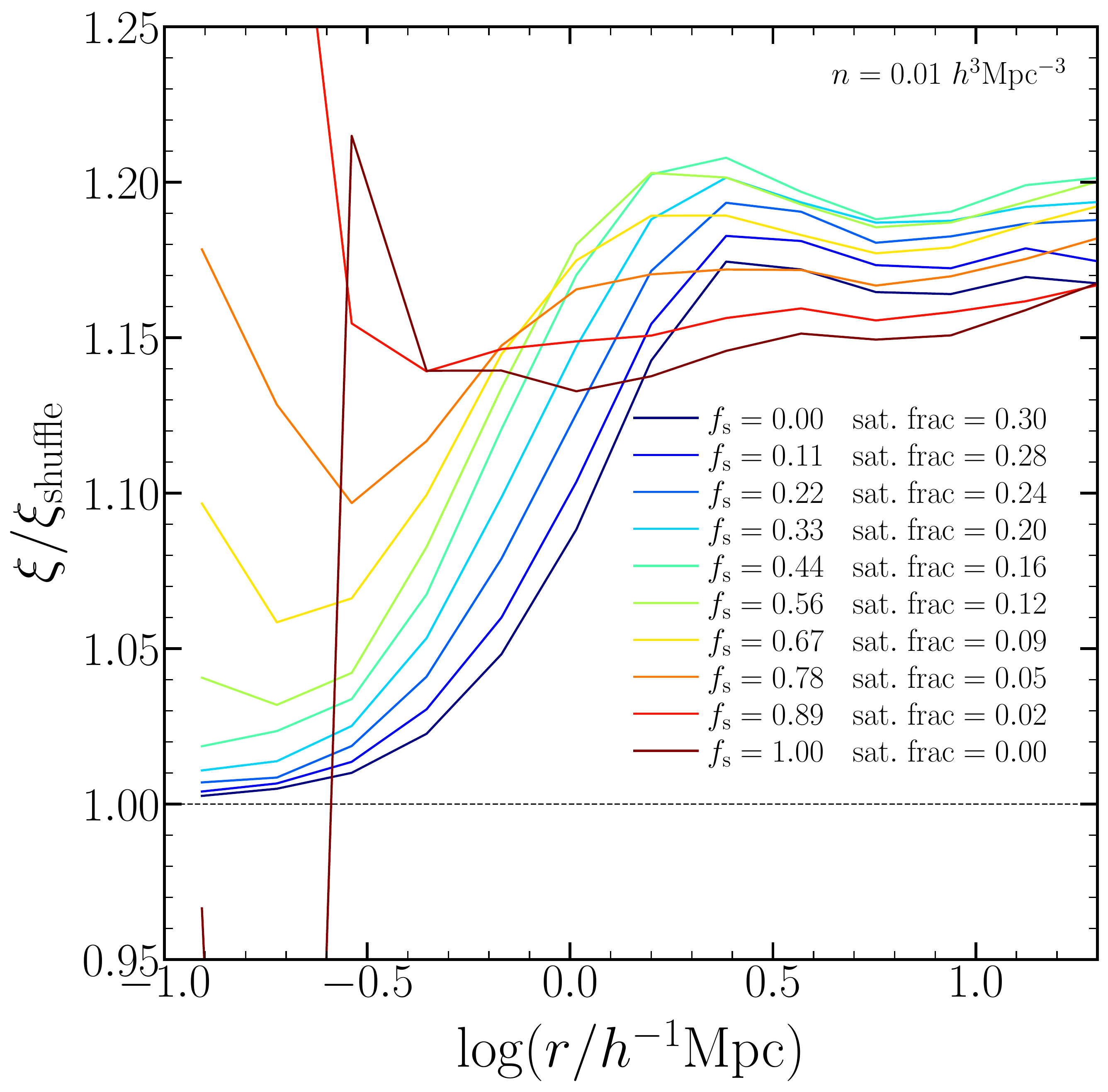}
\includegraphics[width=0.45\textwidth]{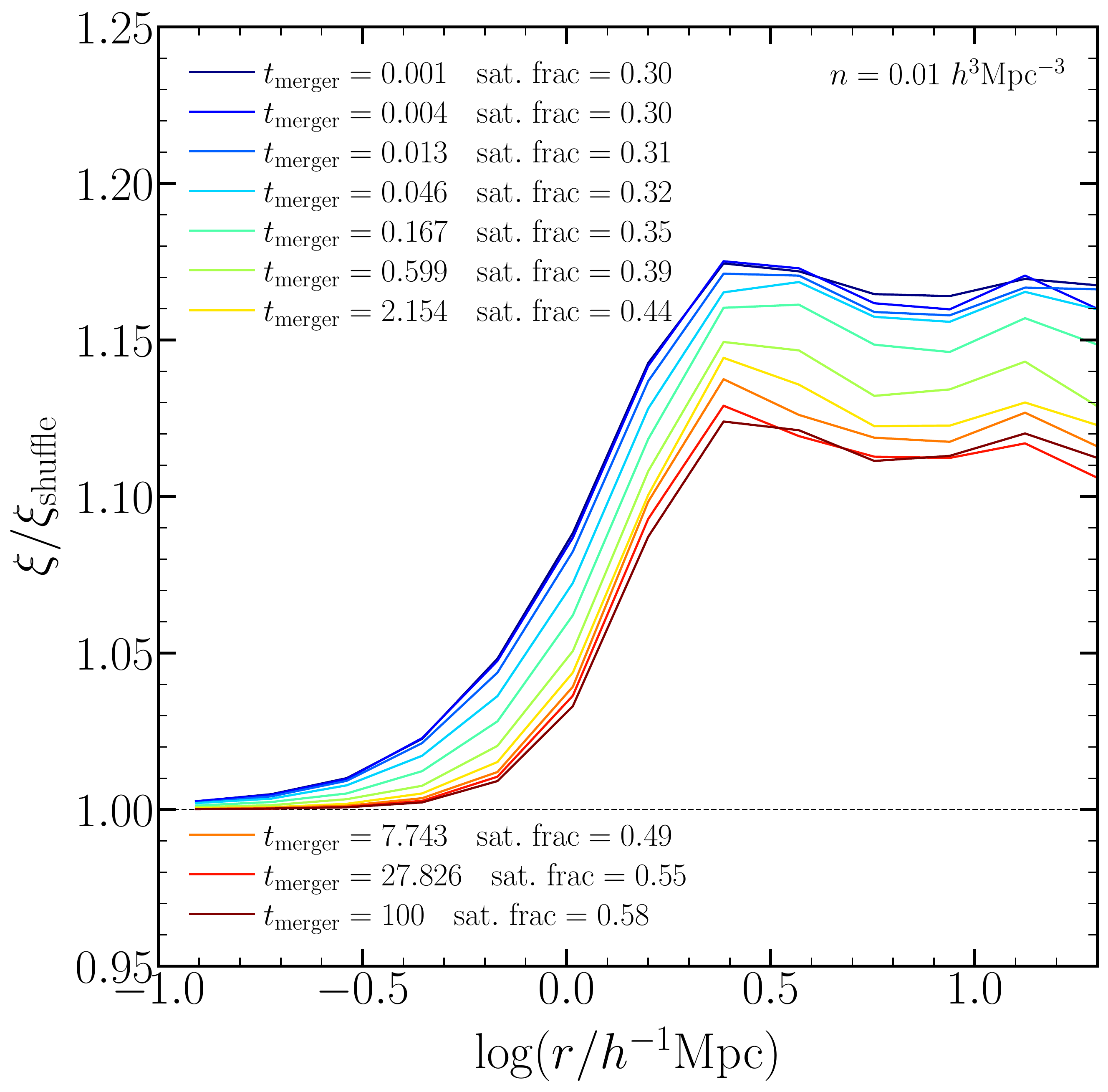}
\caption{(Left panel) The galaxy assembly bias signal expressed as $\xi/\xi_{\rm shuffle}$, for a galaxy sample with a number density of n= 0.01 $\ihMpcC$ and different values of the tidal disruption parameter $f_s$ as labelled. No orphan subhaloes were used to locate galaxies in this run (ie. $\tmerger=0$). A value of $f_s = 0$ means that no satellite galaxy was disrupted, while a value of $f_s = 1$ means all the satellite galaxies were disrupted (so only centrals). The satellite fraction is indicated on the label for each line. The galaxy assembly bias increase with the $f_s$ until it reaches a maximum at $f_s\sim 0.5$, and then decrease.
(Right panel) Similar to the left panel, but varying $\tmerger$, that control the number of orphan galaxies, and assuming no tidal disruption (ie. $f_s=0$). A value of $\tmerger=0$ means no orphan galaxies, while $\tmerger=\infty$ means all possible orphan galaxies are used for constructing the galaxy sample (that is latter selected by stellar mass). As in the left panel, the satellite fraction is indicated for each line in the figure label. Adding satellite galaxies via orphan subhaloes only decrease the level of assembly bias.}
\label{Fig:gab_sat_frac}
\end{figure*}

\section{Galaxy assembly bias at different cosmologies}
\label{sec:GAB}

Hydrodynamical simulations and galaxy formation models predict that the halo occupation of galaxies depends not only on the halo mass but also on other secondary halo properties such as concentration, spin, or age  \citep{Zehavi:2018,Artale:2018}. This effect causes halo assembly bias to propagate and modify the clustering of galaxies. This is known as galaxy assembly bias \citep{Croton:2007}, which can change the  2-point correlation function signal up to 20\% in galaxy formation models such as hydrodynamical simulations \citep[e.g.,][]{ChavesMontero:2016}, semi-analytic models \citep[e.g.,][]{Croton:2007} or even SHAMs \citep[e.g.,][]{C20b}. 

To populate our dark matter simulations, we use the SubHalo Abundance Matching extended model (SHAMe) explained in \S~\ref{sec:SHAMe}. This model can reproduce the galaxy clustering of complex galaxy formation models such as hydrodynamical simulation even when using low-resolution gravity-only simulation. While the assembly bias signal of SHAM models is slightly lower compared to other galaxy models, like hydrodynamical simulations and semi-analytic models \citep{ChavesMontero:2016, C20b}, we expect a consistent behaviour when comparing its amplitude at different cosmologies. We also test our main prediction using a standard SHAM, finding similar results.

To measure the level of assembly bias, we look at the ratio of the 2-point correlation function from our galaxy samples to its shuffled run, following the procedure of \cite{Croton:2007}. The shuffling method consists of exchanging the galaxy population among haloes of similar masses, in particular we use bins of $\Delta \log (M_{\rm h} / h^{-1} \mathrm{M}_\odot) = 0.1$, by construction erasing any dependence of the galaxy occupation on any other halo property besides its mass (ie. no occupancy variation, and by construction, no galaxy assembly bias). Note that we have checked different bin sizes finding comparable results. To reduce the noise from our measurement, we run 4 different SHAMe with different random seeds in each of our simulations. For each of these SHAMe catalogues, we average the signal of 10  independent shuffle runs.

Fig.~\ref{Fig:gab_z0} shows the ratio between the correlation function of the mocks and their shuffling counterpart for two extreme cosmologies at $z=0$. By definition, the ratio between the two correlation functions is equal to the square of the assembly bias signal, $b^2_{\rm gab} = \xi/\xi_{\rm shuffle}$. Similar to Fig.~\ref{Fig:hab_conc} and  Fig.~\ref{Fig:hab_spin}, each panel shows the predictions of two dark matter simulations varying one cosmological parameter at a time as labelled.

Overall, the galaxy assembly bias signal presents low to null differences for different cosmologies. The strongest dependence on cosmology appears when changing $\sig$; in this case, we find differences between 2$\%$-4$\%$ in the total galaxy clustering signal. This difference corresponds up to 35$\%$ variation in $b^2_{\rm gab}$. It is worth pointing out that these differences are small considering the variation in $\sig$. 

These results are similar when we look at the clustering at $z=1$ (Fig.~\ref{Fig:gab_z1}). While the overall galaxy assembly bias signal is lower (in agreement with previous results from SAMs, hydrodynamical simulation and SHAM-like mocks, \citealt{C19,C20b}), there is no major change in the clustering signal as a function of cosmology. We only find small differences when varying $\sig$ and $\wz$. Changing $\sig$ or $\wz$ has a similar effect as changing the redshift of a simulation (e.g., a simulation with low $\sig$ at a given redshift will have a similar galaxy clustering as a simulation with higher $\sig$ at higher redshift). Since galaxy assembly bias evolves with redshift, then it is natural to expect some small differences in galaxy clustering as a function of these parameters. However, we did not find this effect for $\wz$ at $z=0$. This is likely because, at fixed $\sig$, $\wz$ has no impact on the late-time dark matter clustering. We acknowledge that changing $\sig$, $\wz$ or any other parameter is not as simple as the redshift evolution of galaxy clustering; we just want to point out that the change of galaxy assembly bias with cosmology follows the same trends as its evolution with redshift.

One possible explanation for the differences of assembly bias with cosmology (and/or redshift) is the differences in the satellite fraction between the different samples. Galaxy samples at higher redshift, with higher number densities, and/or run in simulations with lower $\sig$ present lower galaxy assembly bias signal. This leads us to think that the satellite fraction could be the main parameter to determine the amplitude of galaxy assembly bias. This would be valid only in models that are capable of reproducing galaxy assembly bias, i.e. HOD models will not show any signal by construction, independently of their satellite fraction. 

Motivated by this, we study the dependence of galaxy assembly bias on satellite fraction in our SHAMe models. We do this for a galaxy sample with a fixed number density of $n=0.01 \ihMpcC$ at $z=0$ by modifying the SHAMe parameters: $f_s$, which controls the disruption of satellite galaxies depending on the amount of mass lost by striping, and $\tmerger$, which controls the amount of orphan galaxies (see \S~\ref{sec:SHAMe} for more details). In Fig.~\ref{Fig:gab_sat_frac} we show the dependence of galaxy assembly bias on $f_s$ (left panel) and $\tmerger$ (right panels). Whenever we modify one of these parameters, we hold fixed the other one to zero. The satellite fraction of each sample is included in the label of the plots. For $f_s$, we notice that there is a non-monotonic relation between the galaxy assembly signal and this parameter. The peak of the signal is reached for a value of $f_s\sim 0.5$, while an only-central sample (i.e. $f_s=1$) or a sample with no subhalo disruption (i.e. $f_s=0$) present the lowest signal. When adding galaxies by populating orphan subhaloes (right panel), we find that the assembly bias signal decreases with the value of $\tmerger$. We also check this assuming other values of $f_s$, finding that the galaxy assembly bias always decreases when adding orphans and that the peak previously found at $f_s\sim 0.5$ is not universal. In Appendix~\ref{sec:appendix} we look into the origin of this evolution, finding that these trends  depend strongly on the redshift of the simulation and the subhalo property used to build the SHAMe.

We conclude from this section that cosmology has a low to null impact on the level of galaxy assembly bias, in a similar way as for halo assembly bias. Only two cosmological parameters have some effect on galaxy assembly bias ($\sigma_8$ and $w_0$), and they only introduce variations of a few percent. These changes are minimal compared to the uncertainty of galaxy formation physics. Studies that aim to look for galaxy assembly bias signal should take this into consideration and use mocks capable of reproducing galaxy assembly bias in the most flexible way possible.

\section{Conclusions}

In this paper, we study the dependence of halo assembly bias and galaxy assembly bias on cosmology. We use a suite of cosmological simulation with different values of $\sig$, $\OmM$, $\Omb$, $\ns$, $\h$, $\Mnu$, $\wz$ \& $\wa$. We first focus on halo assembly bias by selecting haloes according to concentration and spin, motivated by the different halo assembly bias signal induced by these selections \citep{Gao:2007,Mao:2018}. For measuring galaxy assembly bias level, we populate our simulations with the subhalo abundance matching extended technique (SHAMe) presented in \cite{C21a}. This model extends the basic SHAM model by including an orphan and tidal disruption mechanism. This allows us to test how galaxy assembly bias depends on the particular galaxy formation model implemented. 

Here we summarise our more important findings:
\begin{itemize}
  \item We look at the peak height-bias relation for haloes selected by concentration and spin. We find that the halo assembly bias signal is roughly the same for all the cosmologies studied in this work (Fig.~\ref{Fig:hab_conc} and Fig.~\ref{Fig:hab_spin}).
 \item The only cosmological parameter that induces a (small) dependence of the halo assembly bias is $\OmM$. We use large volume simulations ($L_{\rm box}=1440\ \hMpc$) with different values of $\OmM$ to better quantify this effect. We find that changes in $\OmM \in [0.27, 0.36]$ correspond to the differences of only a few percent in bias both when splitting the halo sample according to concentration and spin. This effect only appears for low mass haloes at low redshift (Fig.~\ref{Fig:Rings_HAB}). We checked the robustness of this signal both using higher-resolution simulations and removing splashbacks, reaching to the same conclusions. 
\item We compute the galaxy assembly bias signal for simulations with different cosmologies at $z=0$ and $1$. Again, we find low to null differences between the different cosmologies (Fig.~\ref{Fig:gab_z0} and Fig.~\ref{Fig:gab_z1}). We remind the reader that, since galaxy assembly bias depends on halo assembly bias and the galaxy formation model used, even with an identical halo assembly bias signal, the galaxy assembly signal could have been different if the galaxy formation model had a strong dependence on cosmology. 
\item The largest variations of galaxy assembly bias, albeit of a few percent, appear when varying $\sig$ and $\wz$. This dependence is in similar to the dependence of galaxy assembly bias on redshift found by \cite{C19,C20b}. The effect of these parameters on galaxy clustering is similar to looking at the clustering at a different redshift for a fixed cosmology, which can explain our previous finding.
\item We find some large variations in the galaxy assembly bias signal when changing the satellite fraction in our mocks and the specific way satellites populate haloes (Fig.~\ref{Fig:gab_sat_frac}). We predict that these relationships will probably be different for each galaxy formation model.
\end{itemize}

We conclude that halo and galaxy assembly bias do not have significant dependence on cosmology. Most of the changes in cosmology presented in this work represent at least +/- 10 $\sigma$ around the best cosmological of the  Planck's collaboration \citep{Planck:2018}. On the contrary, we find that the galaxy assembly bias signal has a strong dependence on the particular galaxy formation model used, making the small difference we find for cosmology even less significant. This last finding must be taken into account when designing models to constrain assembly bias from observational data. 

As a final remark, we would like to point out that while galaxy or halo assembly bias shows a negligible dependence on cosmology, this does not mean that having an incorrect cosmology when constraining assembly bias from observation will not affect the analysis, because the overall halo and galaxy clustering still depends on cosmology.

\section*{Data availability}
The data underlying this article will be shared on reasonable request to the corresponding author.

\section*{Acknowledges}

We thanks useful comments from Francisco Germano Maion, Giovanni Aric\`o \& Marcos Pellejero.
The authors acknowledge the support of the ERC-StG number 716151 (BACCO).
The authors also acknowledge the computer resources at MareNostrum and the technical support provided by Barcelona Supercomputing Center (RES-AECT-2019-2-0012 \& RES-AECT-2020-3-0014)
\bibliography{Biblio}

\appendix

\section{Satellite fraction and assembly bias}
\label{sec:appendix}

\begin{figure*}
\includegraphics[width=\textwidth]{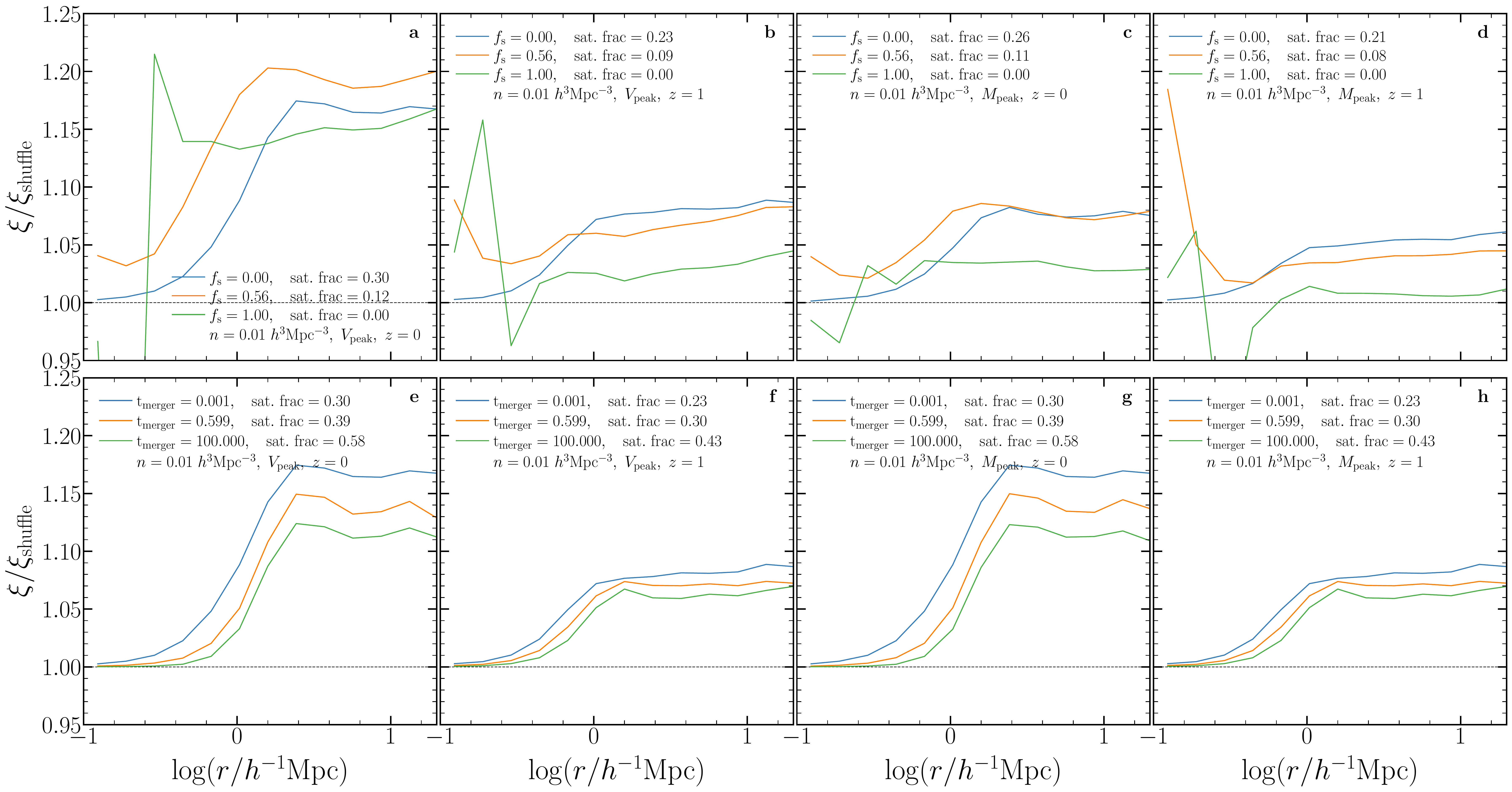}

\caption{The galaxy assembly bias signal expressed as $\xi/\xi_{\rm shuffle}$, for different values of the tidal disruption parameter $f_s$ (panels a, b, c \& d) and $\tmerger$ (panels e, f, g \& h), for a number density of 0.01  $\ihMpcC$ (similar to Fig.~\ref{Fig:gab_sat_frac}). Panels a, c, e \& g show the predictions at $z=0$, while panels b, d, f \& h show the predictions at $z=1$. The SHAMe model used to build the mocks used $V_{\rm peak}$ as their main parameter in panels a, b, e \& f and $M_{\rm peak}$ in panels c, d, g \& h. The satellite fraction of each sample is included in the label of the figure. In each case we show the two extreme values of $f_s$ and $\tmerger$, and an intermediate value; we checked that the trends reported in this figure vary smoothly as we change $f_s$ and $\tmerger$.
}
\label{Fig:gab_sat_frac_all}
\end{figure*}

For a central galaxy only sample (ie. $\Fs=0$ and $\tmerger=0$) we notice in Fig.~\ref{Fig:gab_sat_frac} (see also panels (a) and (e) of Fig.~\ref{Fig:gab_sat_frac_all}) that the level of galaxy assembly bias is positive but lower compare to samples with satellites galaxies add via subhaloes (ie. $\rm f_s>0$,  $\tmerger=0$). By looking at the signal of these two population by separate on fix number density samples, we find that removing centrals increase the level of assembly bias while adding satellite galaxies decreases it.
On the panel (c) of Fig.~\ref{Fig:gab_sat_frac_all} we show a SHAMe model built using $\mpeak$ instead of $\vpeak$ (that should have a low to null level of assembly bias for centrals) and find that adding satellites only decrease the level of galaxy assembly bias, which is in agreement with our previous results. We repeat these tests now at $z=1$ (panels (b), (d) of Fig.~\ref{Fig:gab_sat_frac_all}) finding a stronger decrease of the galaxy assembly bias when we increase the satellite fraction. This is probably because the difference in bias increase as we go to a higher redshifts (or high values of $\nu$).

When looking at galaxies added using orphans subhaloes (panels (e), (f), (g) \& (h) of Fig.~\ref{Fig:gab_sat_frac_all}) we find that in all cases, the galaxy assembly bias decrease with the satellite fraction. We find that our orphan implementation populates preferentially bigger haloes compare to when we add satellites via resolved subhaloes. These subhaloes have a larger negative assembly bias with low concentrated haloes, decreasing the level of galaxy assembly bias in a faster way (valid for $\vpeak$ and $\mpeak$ at $z=0$ and $z=1$)

These findings reinforce our hypothesis that the dependence of galaxy assembly bias is extremely sensitive to the particular galaxy population/formation model used, making all the small dependence with cosmology negligible in comparison.

\bsp	
\label{lastpage}
\end{document}